\documentclass[aps,pra,twocolumn,showpacs,superscriptaddress]{revtex4-1}
\usepackage[latin9]{inputenc}
\usepackage{color}
\usepackage[colorlinks,urlcolor=blue,citecolor=blue,linkcolor=blue]{hyperref}
\usepackage{verbatim}
\usepackage{float}
\usepackage{amsmath}
\usepackage{amssymb}
\usepackage{graphicx}
\usepackage{epstopdf}
\usepackage{esint}
\usepackage{CJK}
\begin{document}
\begin{CJK*}{UTF8}{gbsn} 
\title{Thermometry of a deeply degenerate Fermi gas with a Bose-Einstein condensate}

\author{Rianne S. Lous}
\affiliation{Institut f\"ur Quantenoptik und Quanteninformation (IQOQI), \"Osterreichische Akademie der Wissenschaften, 6020 Innsbruck, Austria}
\affiliation{Institut f\"ur Experimentalphysik, Universit\"at Innsbruck, 6020 Innsbruck, Austria}

\author{Isabella Fritsche}
\affiliation{Institut f\"ur Quantenoptik und Quanteninformation (IQOQI), \"Osterreichische Akademie der Wissenschaften, 6020 Innsbruck, Austria}
\affiliation{Institut f\"ur Experimentalphysik, Universit\"at Innsbruck, 6020 Innsbruck, Austria}

\author{Michael Jag}
\affiliation{Institut f\"ur Quantenoptik und Quanteninformation (IQOQI), \"Osterreichische Akademie der Wissenschaften, 6020 Innsbruck, Austria}
\affiliation{Institut f\"ur Experimentalphysik, Universit\"at Innsbruck, 6020 Innsbruck, Austria}

\author{Bo Huang (黄博)} 
\affiliation{Institut f\"ur Quantenoptik und Quanteninformation (IQOQI), \"Osterreichische Akademie der Wissenschaften, 6020 Innsbruck, Austria}

\author{Rudolf Grimm}
\affiliation{Institut f\"ur Quantenoptik und Quanteninformation (IQOQI), \"Osterreichische Akademie der Wissenschaften, 6020 Innsbruck, Austria}
\affiliation{Institut f\"ur Experimentalphysik, Universit\"at Innsbruck, 6020 Innsbruck, Austria}

\date{\today}

\begin{abstract}
We measure the temperature of a deeply degenerate Fermi gas, by using a weakly interacting sample of heavier bosonic atoms as a probe. This thermometry method relies on the thermalization between the two species and on the determination of the condensate fraction of the bosons. In our experimental implementation, a small sample of $^{41}$K atoms serves as the thermometer for a $^6$Li Fermi sea.  We investigate the evaporative cooling of a  $^6$Li spin mixture in a single-beam optical dipole trap and observe how the condensate fraction of the thermometry atoms depends on the final trap depth.  From the condensate fraction, the temperature can be readily extracted.
We show that the lowest temperature of 5.9(5)\% of the Fermi temperature is obtained, when the decreasing trap depth closely approaches the Fermi energy. To understand the systematic effects that may influence the results, we carefully investigate the role of the number of bosons and the thermalization dynamics between the two species. Our thermometry approach provides a conceptually simple, accurate, and general way to measure the temperature of deeply degenerate Fermi gases. Since the method is independent of the specific interaction conditions within the Fermi gas, it applies to both weakly and strongly interacting Fermi gases. 
\end{abstract}

\maketitle
\end{CJK*}
\section{Introduction}
Since the first demonstration of Fermi degeneracy in an ultracold gas \cite{DeMarco1999oof}, experimental progress has provided unprecedented access to a great wealth of exciting phenomena, as highlighted by the prominent example of a crossover superfluid \cite{Zwerger2012tbb}. The great interest in fermionic quantum gases results from the fact that fermions constitute the elementary building blocks of matter and provide the possibility to investigate various phenomena of strong interactions. The experimental availability of degenerate Fermi gases has led to new insights into intriguing few- and many-body behavior, the many facets of which are studied in a great variety of current experiments. 

The lowest achievable temperature is crucial for the experimental observation of many phenomena. While fermionic superfluidity \cite{Pitaevskii2016bec} is now routinely achieved in many experiments worldwide, other phenomena like anti\-ferromagnetic ordering \cite{Hart2015ooa} require much lower temperatures, which are still very hard to obtain experimentally. In the range of very low temperatures, well below one tenth of the Fermi temperature $T_F$, thermometry becomes increasingly difficult. In deeply degenerate Fermi systems, one faces the general problem that only a small fraction of atoms near the Fermi surface carry the temperature information, which reduces the detection sensitivity for common imaging methods. For strongly interacting systems, the interpretation of density profiles is not straightforward and requires detailed knowledge of the equation of state \cite{Luo2007mot, Nascimbene2010ett, Ku2012rts} to extract temperature information from thermodynamic observables. For the specific case of a unitary Fermi gas with resonant interactions, where thermodynamics follows universal behavior \cite{Ho2004uto}, thermometry is now well established, but not for the general situation of Fermi gases in strongly interacting regimes.

The conceptually most simple way of thermometry is to use a probe in thermal equilibrium with the object under investigation and to rely on a phenomenon with an easily detectable and well-understood temperature dependence. This is the working principle of thermometers in our daily life, where the underlying phenomenon is thermal expansion or temperature-dependent resistivity. We apply the same basic idea to a deeply degenerate Fermi sea, using a small sample of weakly interacting bosonic atoms as a probe, and we rely on the sensitive detection of the condensate fraction.

Our Fermi gas is a spin mixture of deeply degenerate $^6$Li atoms with resonantly tuned interactions, as it is used in many current experiments worldwide. For such a system, temperatures around 10\% of the Fermi temperature $T_F$ or even below have been reported by various groups (see \cite{Yefsah2013hsi,  Lingham2014loo, Burchianti2014eap, Delehaye2015cva, Revelle2016ott} for a few recent examples). Our thermometer is a small sample of bosonic $^{41}$K atoms immersed in the Fermi sea. In addition to the condensate formation serving as the main observable, our system takes advantage of the large mass ratio and the much smaller number of bosons as compared to the fermions. Related thermometry approaches that rely on coupling to a weakly interacting probe component, have been implemented in other Bose-Fermi systems \cite{Roati2002fbq, Ferrier2014amo, Delehaye2015cva, Onofriocat2016}, in population-imbalanced spin mixtures \cite{Zwierlein2006doo}, and in a Fermi-Fermi mixture \cite{Spiegelhalder2009cso}, but without combining all these three advantages. For our system, the critical temperature for Bose-Einstein condensation (BEC) corresponds to about  0.1\,$T_F$,  which makes the condensate fraction a sensitive and accurate probe right in the temperature range of main interest for deep cooling.
 
In this paper, we present a thorough experimental investigation of Fermi gas thermometry using a bosonic species. In Sec.~\ref{sec:basicidea}, we discuss the basic principle of thermometry for a Fermi-Bose system in general and for the particular case of our mixture of $^6$Li and $^{41}$K. We then, in Sec.~\ref{sec:implement}, describe the experimental procedures of preparation, cooling, trapping, and detection. In Sec.~\ref{sec:results}, we present the main experimental results on deep cooling of the $^6$Li spin mixture, as probed by the $^{41}$K BEC.


\section{Bosons as a Fermi gas thermometer}
\label{sec:basicidea}

Here, we first discuss the basic idea of our thermometry approach in general terms, before we turn our attention to the specific case of $^{41}$K bosons in a $^6$Li Fermi sea.

\subsection{Basic idea}
\label{ssec:general}

\begin{figure}
\includegraphics[clip,width=0.95\columnwidth]{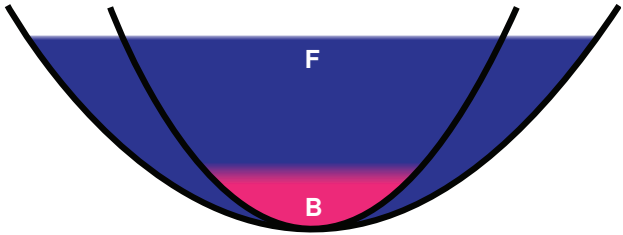}
\caption{(Color online) Basic idea of the thermometry. A small sample of bosonic atoms (B) is immersed in a large, deeply degenerate sea of fermions (F) under thermal equilibrium conditions. The harmonic trapping potentials (solid lines) are different for both species, depending on the particular trapping configurations used. The temperature is derived from the condensate fraction.}
\label{fig:basicidea}
\end{figure}

The basic idea of our thermometry approach is illustrated in Fig.~\ref{fig:basicidea}. We assume that both harmonically trapped species are in sufficient thermal contact with each other to establish a thermal equilibrium with a common temperature $T$. The main observable is the condensate fraction $\beta$  of the bosonic cloud, from which $T$ can be derived. 

To obtain the temperature $T$ of the two-component system in relation to the Fermi temperature $T_F$, we start with the identity $T/T_F = (T/T_c) \times (T_c/T_F)$, where $T_c$ is the critical temperature for BEC. The first factor, $T/T_c$, can readily be obtained from the condensate fraction of the bosonic component through the well-known relation
\begin{equation}
\frac{T}{T_c}  =  (1 - \beta)^{1/3}  \, .
\label{eq:beta}
\end{equation}
For calculating $T_c/T_F$ we use the textbook formulas

\begin{eqnarray}
k_B T_c  & =  0.940  \, \hbar \omega_B \, N_B^{1/3} \, ,  \label{eq:TTc} \\ 
k_B T_F & =  1.817  \, \hbar \omega_F \, N_F^{1/3}  \, , \label{eq:TTF}
\end{eqnarray}
where $N_B$ and $N_F$ represent the number of trapped bosons and fermions, and $\omega_B$ and $\omega_F$ are the respective geometrically averaged trap frequencies. Note that Eqs.~(\ref{eq:beta})and (\ref{eq:TTc}) are strictly valid only for non-interacting systems in the thermodynamic limit. In practice, the finite sample size and interaction effects may lead to corrections~\cite{Giorgini1996cfa}.

By combining Eqs.~(\ref{eq:beta})-(\ref{eq:TTF}) we arrive at the central equation that underlies our thermometry approach,
\begin{equation}
\frac{T}{T_F} = 0.518 \, (1 - \beta)^{1/3} \,\, \frac{\omega_B}{\omega_F} \left( \frac{N_B}{N_F}\right)^{1/3} \, .
\label{eq:basicidea}
\end{equation}
In an experiment, the ratio of the trap frequencies, $\omega_B/\omega_F$, will be determined by the specific properties of the two different components and the particular trap configuration used for the experimental realization.  

Equation~(\ref{eq:basicidea}) highlights the conditions for optimized thermometry in the deeply degenerate regime. A small ratio of the trap frequencies, $\omega_B/\omega_F$, is highly desirable. This favors heavy bosons in combination with light fermions. The number ratio $N_B/N_F$ enters with its third root, which shows that a very large number imbalance ($N_B \ll N_F$) is required to take real advantage of this factor. In this case, the bosons can be considered as impurities in the large Fermi sea.

\subsection{Case of the $^6$Li-$^{41}$K mixture}
\label{ssec:LiK}

We now turn our attention to the specific situation of bosonic  $^{41}$K atoms in a Fermi sea of $^6$Li atoms. The mixture \cite{Wu2011sii, Yao2016ooc}  exhibits favorable properties for our purpose. The interspecies interaction is moderate, with a background scattering length of about +60\,$a_0$ \cite{Hannapriv}, where $a_0$ is Bohr's radius. This is large enough to provide a sufficient cross section for thermalization on a realistic experimental time scale, but weak enough to avoid effects of strong interactions, such as a mutual repulsion or attraction or fast three-body decay. 

We consider a hybrid trapping scheme, as realized in our experiment, where the atoms are confined radially by an infrared laser beam and axially by a curved magnetic field (see Sec.~\ref{ssec:trap}), under conditions ensuring that the trap frequency ratio for the two species is not changed by the gravitational sag (see Appendix \ref{app:magic}). For such a trap, in a harmonic approximation, the frequency ratio in Eq.~(\ref{eq:basicidea}) can be expressed as
\begin{equation}
\frac{\omega_B}{\omega_F} = 
\left( \frac{m_{\rm K}}{m_{\rm Li}} \right)^{-1/2} \left( \frac{\alpha_{\rm K}}{\alpha_{\rm Li}} \right)^{1/3} \left( \frac{\mu_{\rm K}}{\mu_{\rm Li}} \right)^{1/6} \, ,
\label{eq:ratios}
\end{equation}
For our experimental situation (Sec.~\ref{ssec:trap}), the mass ratio is ${m_{\rm K}}/{m_{\rm Li}} = 6.810$, the ratio of optical polarizabilities is ${\alpha_{\rm K}}/{\alpha_{\rm Li}} = 2.209$  \cite{Tang2010ddp, Safronova2013mwf}, and  
the ratio of magnetic moments is ${\mu_{\rm K}}/{\mu_{\rm Li}} = 0.999$. With these accurately known numbers, Eqs.~(\ref{eq:basicidea}) and (\ref{eq:ratios}) yield
\begin{equation}
\frac{T}{T_F} = 0.258 \, (1 - \beta)^{1/3} \, \left( \frac{N_B}{N_F}\right)^{1/3} \, ,
\label{eq:basicideax}
\end{equation}
which we will use for extracting  $T/T_F$ from our experimental data, as described in the following sections.

The dynamical range of our thermometry approach as applied to the $^{41}$K-$^6$Li mixture can now be illustrated by a numerical example, based on typical experimental conditions. We assume $N_B/N_F = 1/30$ and possible measurements of the condensate fraction in the range $0 \le \beta \lesssim 0.95$. According to Eq.~(\ref{eq:basicideax}), this corresponds to a temperature range of $0.03 \lesssim T/T_F \lesssim 0.08$, right in the interesting regime for state-of-the art experiments in the deeply degenerate Fermi gases.

\section{Experimental procedures}
\label{sec:implement}

In this section, we present our general experimental procedures applied to a Fermi-Bose mixture of $^6$Li and $^{41}$K. In Sec.~\ref{ssec:prepare}, we give an overview of the main preparation steps. In Sec.~\ref{ssec:trap}, we present in detail the optical dipole trap used in the final stage of deep evaporative cooling. In Sec.~\ref{ssec:detect}, we discuss the main detection schemes.

\subsection{Preparation of the $^6$Li-$^{41}$K mixture}
\label{ssec:prepare}

The mixture is prepared in an all-optical cooling and hybrid trapping approach, very similar to the one described in detail in Ref.~\cite{Spiegelhalder2010aop} and applied in various previous experiments on the mixture of $^6$Li and $^{40}$K atoms (see, e.g.,\ Refs.~\cite{Spiegelhalder2009cso, Trenkwalder2011heo, Kohstall2012mac, Jag2014ooa, Cetina2016umb}). A spin mixture of $^6$Li atoms in the lowest two sublevels of the electronic ground state is evaporatively cooled close to a Feshbach resonance \cite{Ohara2002ooa, Bourdel2003mot, Jochim2003bec} and serves as the agent to sympathetically cool a K minority component. For the whole cooling process, we found that it makes no difference whether the fermionic $^{40}$K or the bosonic $^{41}$K isotope is present, if we avoid any interspecies scattering resonances and rely on the background interaction with the $^6$Li cooling agent, being about the same for both K isotopes.

The preparation process involves  a spin relaxation stage, which we employ to prepare a single K spin state. Here, the parameters differ from our previous work on $^{40}$K  \cite{Spiegelhalder2010aop}.  For $^{41}$K, the initial laser cooling stage provides an unpolarized sample in the three magnetic sublevels ($m_F = -1, 0, 1$) of the lowest hyperfine level ($F=1$).  We found \cite{Lous2017PhD} that spin-exchange collisions with $^6$Li atoms in the second-lowest sublevel can efficiently produce a  polarized $^{41}$K sample in the $m_F = -1$ state, which is the third-lowest Zeeman sublevel. The spin relaxation is performed near a magnetic field of 200\,G, where the process appears to be resonantly enhanced. This stage has a duration of about 500\,ms and is implemented right after loading the optical dipole trap, when the temperature is still rather high (few 100\,$\mu$K).  To remove a residual population of K in the  $m_F = 0$ state (typically 15\%), we apply a resonant laser pulse right before the final evaporation stage to push those atoms out of the trap. It is interesting to note that, without applying the spin cleaning, the evaporation process leads to a spinor condensate \cite{stamperkurn2013sbg} with clear signatures of immiscibility \cite{Liu2016sgd}. The $s$-wave background interaction between the bosons is relatively weak (intraspecies scattering length of +60\,$a_0$ \cite{DErrico2007fri, Patel2014frm}), which makes the condensate very stable against three-body decay.

\subsection{Trap for deep evaporative cooling}
\label{ssec:trap}

\begin{figure}
\includegraphics[clip,width=0.95\columnwidth]{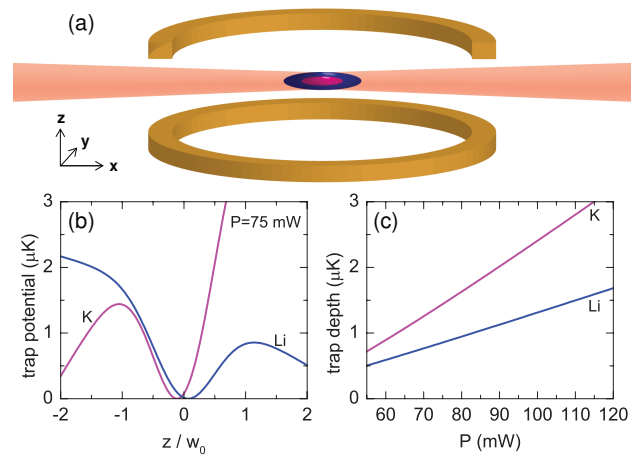}
\caption{(Color online) Trapping scheme in the final stage of evaporative cooling. (a) A single infrared laser beam for radial trapping ($y,z$ directions) is used in combination with a magnetic field (coil setup schematically shown). The magnetic field serves multiple purposes, providing the bias field for Feshbach tuning, a curvature for axial trapping ($x$ direction), and a vertical levitation gradient. (b) The vertical potentials $U^i_{\rm tot}(z)$ resulting from Eq.~(\ref{eq:Utot}) for both Li (blue) and K (magenta) are shown for a typical laser power of $P = 75$\,mW. For illustrative purposes, we have introduced species-dependent offsets to shift the potential minima to zero. (c) The trap depths $U^i_{\rm trap}$ depend on the laser power $P$, with the K trap being always deeper than the Li trap.}
\label{fig:trap}
\end{figure}

The whole evaporation process takes place in several stages  \cite{Spiegelhalder2010aop} within a total time of 12\,s. Here, we focus on the final stage, where the power of a single laser beam is ramped down exponentially within 5\,s, from an initial value of 440\,mW to a final value in the range between 110 and 45\,mW. Then, a hold time of 10\,s is introduced to ensure full thermalization, before the two species are finally detected; see Sec.~\ref{ssec:detect}. As in our previous work \cite{Spiegelhalder2009cso, Trenkwalder2011heo, Kohstall2012mac, Jag2014ooa, Cetina2016umb}, the magnetic bias field of 1180\,G is applied for standard Feshbach tuning of the interaction between the two $^6$Li spin components. This leads to a large negative $s$-wave scattering length of $a =  - 2900\,a_0$  \cite{Zurn2013pco},
and thus facilitates highly efficient evaporative cooling with very low inelastic losses. We note that, because of the absence of any significant losses, the number of K atoms stays essentially the same during the whole evaporative cooling process.

We hold the spin mixture of $^6$Li together with the single spin state of $^{41}$K in a hybrid trap \cite{Jochim2003bec} as illustrated in Fig.~\ref{fig:trap}(a). Here, the radial confinement ($y, z$ directions) is provided by a single 1064-nm laser beam and the axial confinement ($x$ direction) results from the curvature of the applied magnetic field. In the vertical direction, gravity also comes into play and decreases the trap depth, which influences both species differently. We apply an additional magnetic levitation field to compensate for the latter effect. 
The levitation potential is given by
\begin{equation}
U_{\rm lev}(z) = - \mu_B B' z \, ,
\end{equation}
where $\mu_B$ is Bohr's magneton and $B'$ represents the vertical gradient of the magnetic field. Note that for our high bias magnetic field of 1180\,G the levitation potential is essentially the same for both species, since the magnetic moments of both species are within 0.1\% close to $\mu_B$. We use a gradient of $2.5(2)$\,G/cm, for which we obtain $\mu_B B' / m_{\rm K} g = 0.34(3)$, {\it i.e.}\ we realize a partial levitation of the K atoms by compensating one-third of the effect of gravity (gravitational acceleration $g$). For Li, we obtain $\mu_B B' / m_{\rm Li} g = 2.36(20)$, which means a strong overlevitation. These conditions are close to a `magic' levitation condition, where the combined tilt effect of gravity and levitation on the trap depth is the same for both species; see Appendix \ref{app:magic} for a detailed description.

For both species ($i = $Li, K), the total potential along the vertical direction in the trap center can be written as
\begin{eqnarray}
U_{\rm tot}^{i}(z)  & = & - U^{i}_{\rm opt} \, \exp(-2z^2/w^2) \nonumber \\
& & + (m_{i}g - \mu_B B') z - \frac{1}{2} \mu_B B'' z^2 \, ,
\label{eq:Utot}
\end{eqnarray}
where $U^{i}_{\rm opt}$ is the optical potential depth and $w$ is the waist of the single optical beam. 
The combined effect of gravity and magnetic levitation is represented by the term linear in $z$. The quadratic term describes a weak magnetic antitrapping effect, resulting from the negative curvature of the magnetic field. In the saddle-potential of our configuration [Fig.~\ref{fig:trap}(a)], the curvature along the $z$ axis is two times larger and of opposite sign as compared to the curvature along the $x$ axis, the latter determining the axial magnetic confinement. Therefore, the curvature $B''$ is related to the axial trapping frequency $\omega^i_x$ by the formula $\mu_B B'' = 2 m_i (\omega^i_x)^2$.

The vertical trap potentials are shown in Fig.~\ref{fig:trap}(b) for both Li and K atoms under typical conditions of our experiment ($P = 75\,$mW). This clearly illustrates the different optical potentials and the effect of the opposite tilt on both Li and K.  The tilt and the curvature substantially reduces the total trap depths $U^i_{\rm trap}$ to values below the respective depths of the optical potentials ($U^i_{\rm trap} < U^i_{\rm opt}$).

 Figure~\ref{fig:trap}(c) illustrates the dependence of the trap depths $U^i_{\rm trap}$ on the laser power in the range relevant for our final evaporative cooling stage. It is important to note that $U^{\rm K}_{\rm trap} > U^{\rm Li}_{\rm trap}$ is always fulfilled. The effect of the magnetic levitation ensures that evaporative cooling removes Li atoms, but leaves all K atoms in the trap. This is essential for our interpretation of the cooling process, where Li acts as the cooling agent and K is cooled sympathetically via collisions with Li and not directly. 

We characterize the trap by measuring the frequencies of radial and axial sloshing oscillations of both the confined species. 
For the radial trap frequencies of Li and K, we find
\begin{subequations}
\begin{eqnarray}
\omega^{\rm Li}_r  & = 2\pi \times 37.6(5){\rm Hz} \times \sqrt{ P / {\rm mW} } \, , \label{eq:fradLi} \\
\omega^{\rm K}_r  & = 2\pi \times 21.0(6){\rm Hz} \times \sqrt{ P / {\rm mW} } \, ,
\label{eq:fradK}
\end{eqnarray}
\end{subequations}
where $P$ is the power of the trapping beam. The measured frequency ratio $\omega^{\rm Li}_r / \omega^{\rm K}_r = 1.79(6)$ is consistent with the more accurate value of 1.756 as calculated from the dynamic polarizabilities \cite{Tang2010ddp, Safronova2013mwf} and the mass ratio. For the single-beam optical dipole trap, assuming a Gaussian laser beam profile, we then obtain \cite{Grimm2000odt} the waist $w = 44.3\,\mu$m and the optical potential depths
\begin{subequations}
\begin{eqnarray}
U^{\rm Li}_{\rm opt}  / (k_B \times {\rm nK}) =  19.8(3)\, P / \rm mW \, ,\\
U^{\rm K}_{\rm opt}  / (k_B \times {\rm nK})  =  43.7(6)\, P / \rm mW \, .
\label{eq:U0}
\end{eqnarray}
\end{subequations}

For the axial frequencies, characterizing the magnetic confinement, we obtain 
\begin{equation}
\omega^{\rm Li}_x = 2.61\,\omega^{\rm K}_x = 2 \pi \times 25.6(1)\,{\rm Hz} \, .
\label{eq:fax}
\end{equation} 
We note that, for the trap frequencies, the optical contribution to the axial trapping and magnetic effects on the radial confinement remain negligibly small. Furthermore, the levitation field that counteracts gravity leaves the oscillation frequencies at the bottom of the trap essentially the same \cite{Hung2008aec}, in spite of its substantial effect on the trap depths. This ensures that the frequencies according to Eqs.~(\ref{eq:fradLi}) and (\ref{eq:fradK}) remain a very good approximation for all our experimental conditions.

\subsection{Detection}
\label{ssec:detect}

For detection of the two species we apply standard time-of-flight absorption imaging, realized with probe beams propagating along the $z$ axis. From images of the $^6$Li cloud, we selectively determine the number $N_F$ of fermionic atoms in each of the two lowest spin states with relative uncertainties of about 15\% \cite{Cetina2015doi}. For $^{41}$K, we detect the number $N_B$ of bosonic atoms in the third-to-lowest spin state with an estimated relative uncertainty of 15\%. From the images of the bosons, we also extract the condensate fraction $\beta$, which is the quantity of main interest for our thermometry approach. 

Time-of-flight absorption imaging of the expanding $^{41}$K component can, in principle, be implemented by turning off the trapping laser beam and letting the cloud expand in the same magnetic field configuration as it is used for evaporative cooling. However, in such a simple scheme, the magnetic field curvature causes a focusing effect \cite{Donley2001doc} in the $x,y$ plane (oscillation frequency $\sim$10\,Hz), which occurs right in the time interval of main interest for the imaging. For analyzing the ballistic expansion of the thermal cloud, it is rather straightforward to take the focusing effect into account \cite{Ketterle2007mpa}, so that the temperature can be readily extracted. For the condensed part, however, the focusing effect leads to an increase of the density and the optical depth of the cloud, which makes a determination of the condensate fraction problematic. 

We employ a modified scheme for time-of-flight absorption imaging, where we adiabatically transform our hybrid trap into a purely optical one, before the cloud is released into free space. To prevent any effect of interspecies interaction in the transfer stage, we remove the Li atoms before the transfer into the crossed-beam trap by smoothly applying a short stage with a magnetic gradient of about 8\,G/cm, which levitates the K cloud and spills all Li atoms out of trap. Then we slowly ramp up a second trapping beam, which has a fixed final power of $P'=70$\,mW, and a waist of $\sim$\,$50\,\mu$m and crosses the first beam under an angle of $16^{\circ}$ \cite{Cetina2015doi}. The magnetic field is simultaneously changed to a homogenous configuration without curvature, but with the same bias field. The potential of the resulting crossed-beam dipole trap is similar to the hybrid trap of the evaporation stage and the transfer is realized over a rather long time of 4\,s, which ensures adiabaticity of the process. The transfer into the detection trap, being somewhat tighter than the cooling trap, implies a moderate adiabatic compression. This increases the temperature by a factor of roughly 1.5, as easily obtained from the ratio of the trap frequencies \footnote{At an intermediate power of $P=75$\,mW, the geometrically averaged trap frequency increases from 69 to 109\,Hz. The frequency ratio depends on  the value of $P$, but quite weakly. We have carried out measurements on the trap frequencies in the crossed-beam detection trap, from which we determine the change in trap frequencies with an accuracy of about a few percent.}. This factor is taken into account when we determine the temperature of the thermal component from the temperature of the expanding cloud. To image the expanding cloud after time of flight, we apply a levitation field that counteracts gravity and facilitates long observation times up to 45\,ms.

We have performed several tests on the performance of our detection scheme. We have carefully checked that the adiabatic transfer stage does not lead to any detectable loss of K atoms and that its influence on the condensate fraction remains negligibly small. 

\section{Cooling and thermometry results}
\label{sec:results}

In this section, we present our experimental results. We focus on the final stage of the deep evaporative cooling process, where the lowest temperatures are achieved. 
In Sec.~\ref{ssec:spill}, we consider the fermionic $^6$Li component only and identify the conditions where cooling crosses over into spilling of the Fermi sea. 
In Sec.~\ref{ssec:equil}, we turn our attention to the bosonic $^{41}$K component and present measurements of the condensate fraction and the temperature, 
which allows us to determine $T/T_F$ for the Fermi gas. In Sec.~\ref{ssec:therm}, we investigate the interspecies thermalization process, justifying the assumption of interspecies thermalization. In Sec.~\ref{ssec:discuss}, we finally discuss the performance of our thermometry scheme in terms of measurement uncertainties and systematical effects.

\subsection{Crossover from evaporation to spilling}
\label{ssec:spill}

In the final stage of evaporative cooling, when the laser power is reduced to very low values, a crossover between two regimes takes place \cite{Jochim2003bec}. Above a certain threshold, the continuous reduction of the trap power removes thermal atoms with some excess energy above the Fermi energy level, which efficiently cools down the sample. Then a threshold is reached where the Fermi energy level in the shallow trap reaches the trap depth. Below that threshold, the atoms are spilled out of the trap. We identify this crossover by measuring the number of $^6$Li atoms remaining in the trap as a function of the final trap power at the end of the evaporation ramp.

\begin{figure}
\includegraphics[clip,width=1.00\columnwidth]{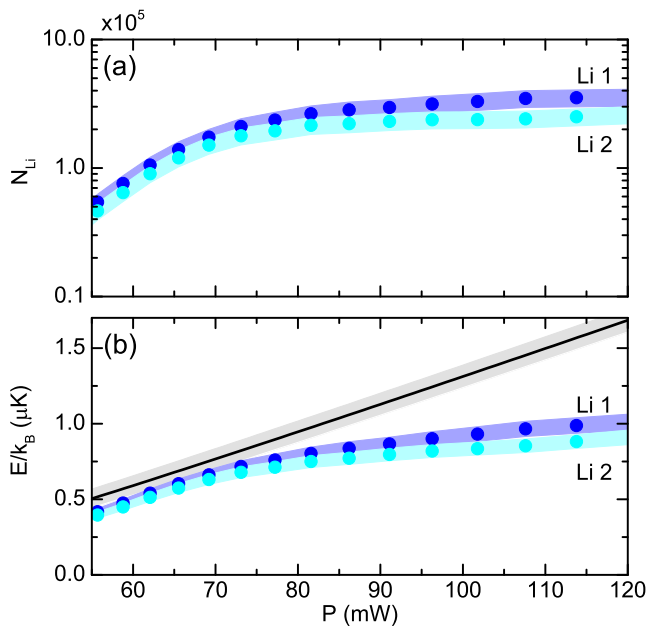}
\caption{(Color online) Crossover from the cooling to the spilling regime in deep evaporative cooling of $^6$Li.  
In (a), we show the measured dependence of the atom number in both spin states as a function of the laser power $P$, which decreases during the evaporation ramp. Here the labels Li\,1 and Li\,2 refer to the lowest and second-to-lowest spin state of Li, respectively. The systematic calibration uncertainty in the number determination ($\pm15$\%) is indicated by the shaded error band.
In (b), we plot the corresponding behavior of the Fermi energy $E_F$ and compare it with the decreasing trap depth $U^{\rm Li}_{\rm trap}$ (solid line). The shaded region indicates a systematic uncertainty in the trap depth resulting from the determination of the levitation gradient, which we consider as the dominant error source for $E_F$.}
\label{fig:evapLi}
\end{figure}

Figure \ref{fig:evapLi} shows our observations for a final trap power $P$ between 45 and 110\,mW. The crossover between the two different regimes can be clearly seen in  the behavior of both the atom numbers (a) and the resulting Fermi energies (b). The results reveal a change between $70$ and $80\,$mW, which marks the crossover into the spilling regime. This interpretation is further confirmed by the behavior of the trap depth, as calculated from Eq.~(\ref{eq:Utot}). Below a power of about 70\,mW, the corresponding solid line in (b) gets very close to the data points and shows essentially the same slope \footnote{The Fermi energy is calculated in the harmonic approximation. We estimate that the anharmonicity of the trap leads to an error on the order of 5\%.}. It is also interesting to note that the spilling effect removes a small initial imbalance in the population of both spin states.

As we will see in Sec.~\ref{ssec:equil}, the deepest cooling takes place in the discussed crossover regime. We therefore summarize the relevant experimental parameters at $P = 75$\,mW, where we have $N_F = 2.0 \times 10^5$ atoms per spin state in a trap with an average frequency $\omega_F = 2 \pi \times 140\,$Hz. This results in a Fermi energy of $E_F = k_B \times 710$\,nK, corresponding to a peak number density of $n_F = 1.3 \times 10^{12}$\,cm$^{-3}$ per spin state and a Fermi wavenumber of $k_F = 1/(4500\,a_0)$.  

The interaction in the spin mixture \cite{Zwerger2012tbb} is characterized by the parameter $1/(k_F a) \approx -1.6$, which shows that our gas is not in the strongly interacting regime as defined by $|1/(k_F a)| < 1$, but also not far away from it. The attraction in the gas can be estimated \cite{Navon2010teo} to have $\sim$10\% effect on the chemical potential and the number density as compared to the interaction-free values. We point out that this does not play any role for our thermometry approach, because we probe the temperature with another species. This is in contrast to temperature measurements that are based on the size and shape of the trapped cloud. The latter require knowledge of the temperature-dependent equation of state \cite{Ku2012rts} for the particular interaction conditions. 

\subsection{Condensate fraction and equilibrium temperatures}
\label{ssec:equil}


Here, we first present our measurements of the condensate fraction, from which we derive the relative temperature $T/T_F$. Then we compare these results with direct temperature measurements of the thermal fraction of the bosons, and we finally investigate how the number of bosonic atoms affects our results.

\begin{figure}
\includegraphics[clip,width=1.00\columnwidth]{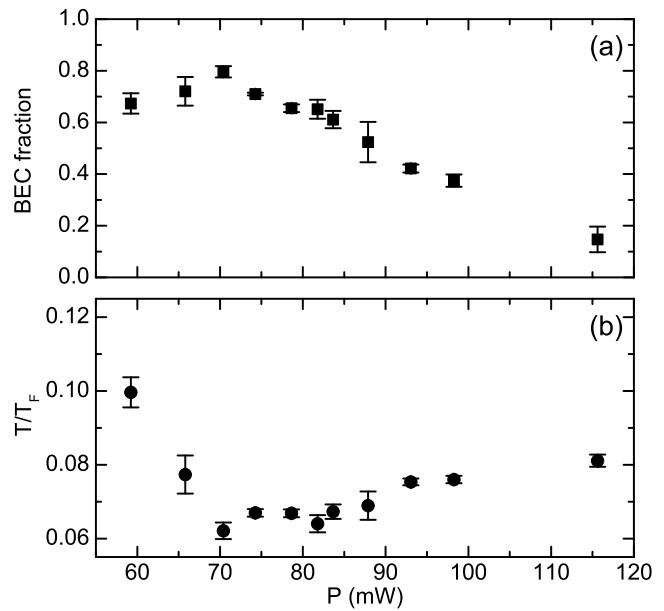}
\caption{Fermi gas thermometry based on partially condensed bosons. (a) The measured condensate fraction $\beta$ is shown as a function of the final power $P$ of the evaporation ramp. Here the small error bars (most of them smaller than the symbol size) reflect the uncertainties of bimodal fits to time-of-flight images. In (b), we show the corresponding results for the relative temperature $T/T_F$. Here the error bars reflect the total statistical uncertainties from fitting the condensate fraction and the atom numbers, but not the calibration uncertainties in the atom numbers. The latter result in an additional systematic scaling uncertainty of $\pm7$\%.}
\label{fig:becfrac}
\end{figure}


Figure~\ref{fig:becfrac}(a) shows the BEC fraction $\beta$, measured as a function of the final power $P$ of the evaporation ramp. Each data point is the mean derived from images taken at seven different times of flight (12 to 24\,ms), with the corresponding standard error of the mean. The total number of bosonic $^{41}$K atoms is $N_B \approx 1.3\times 10^4$, independent of $P$. 
We locate the condensation threshold somewhere near $125$\,mW and, with decreasing power, we observe a steady increase of the condensate fraction until a maximum of $\beta \approx 0.8$ is reached near 75\,mW. The conditions of the Fermi sea of $^6$Li atoms are exactly the ones already described in the preceding section.

Using Eq.~(\ref{eq:basicideax}) and applying small finite-size and interaction corrections to the critical temperature~\cite{Giorgini1996cfa}, we derive the relative temperature $T/T_F$ for the degenerate Fermi gas~\footnote{Finite-size effects and interaction effects lead to small downshifts of $T_c$. To derive the temperature from the condensate fraction, for the sake of simplicity, we use Eq. (1) with corrections to $T_c$ from~\cite{Giorgini1996cfa}. 
 Even at our smallest atom numbers, the temperature corrections stay well below 10\%. Interaction corrections in our largest clouds stay below 2\%.}. 
The results are shown in Fig.~\ref{fig:becfrac}(b). We observe lowest values of $T/T_F \approx 0.07$  for $P$ between 70 and 85\,mW. This power range corresponds to what we have identified before as the crossover regime between evaporative cooling and spilling. In the spilling regime, we see an increase in the relative temperature, due to a fast spilling of the Li atoms. We conclude that the deepest degeneracy of the Fermi gas is achieved when the evaporation is stopped just before the onset of spilling.

\begin{figure}
\includegraphics[clip,width=1.00\columnwidth]{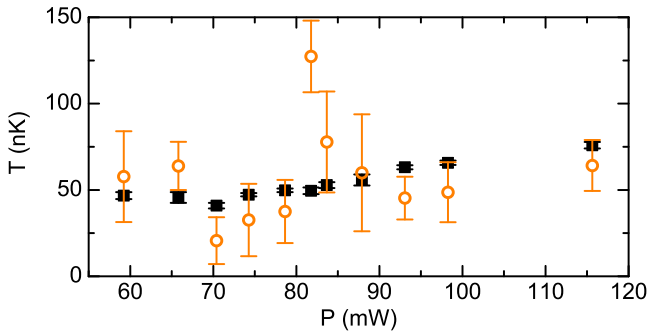}
\caption{Comparison of two methods to obtain the temperature from time-of-flight images. The filled symbols represent the temperatures determined from the condensate fraction [see data in Fig.~\ref{fig:becfrac}(a)] together with the total number of bosons and the separately measured trap frequency. The open symbols represent the temperatures that result from the expansion of the thermal component of the bosonic cloud. For the closed symbols, most of the errors derived are smaller than the symbol size. These errors represent the statistical uncertainties as derived from measurements at seven different expansion times. For the open symbols, the error bars are the uncertainties from fits to the expansion dynamics.}
\label{fig:Tcompare}
\end{figure}


Figure \ref{fig:Tcompare} displays the absolute temperature $T$ derived according to Eqs.~(\ref{eq:beta}) and (\ref{eq:TTc}) from the BEC fraction data already presented in Fig.~\ref{fig:becfrac}(a). We compare these results with the temperature of the thermal component, which we extract from the same time-of-flight images by fitting the expansion dynamics. The comparison shows that both methods provide consistent results, but it also reveals much smaller statistical uncertainties (see error bars) for the first method. This observation highlights an important advantage for accurate thermometry of our method that is based on the determination of the condensate fraction.

\begin{figure}
\includegraphics[clip,width=1.00\columnwidth]{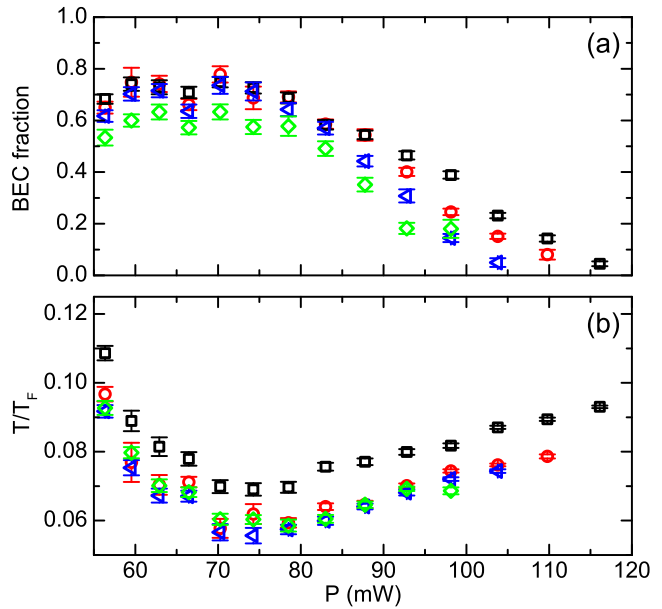}
\caption{Influence of the number of bosons on (a) the condensate fraction $\beta$ and (b) the resulting relative temperature $T/T_F$. Four different data sets are shown, with different numbers of bosonic K atoms: black squares, $N_B = 1.5(1)\times 10^4$; red circles, $1.2(1)\times 10^4$; blue triangles, $0.9(1)\times 10^4$; green diamonds, $0.76(6)\times 10^4$. The error bars represent the statistical uncertainties as derived from the fit errors of the condensate fraction.}
\label{fig:varyNB}
\end{figure}


In an additional set of experiments, we have addressed the question of whether the presence of the $^{41}$K bosons has an influence on the cooling of the Fermi gas. We therefore reduced the number of K atoms from about 15,000 (similar to Fig.~\ref{fig:becfrac}) down to about 7500. Here, for the sake of shorter data acquisition time, we applied a simpler, but somewhat less accurate detection scheme than before \footnote{In Fig.~\ref{fig:varyNB}, we have used a single time of flight of 22\,ms to reduce the total measurement time. This  method may be somewhat less accurate, but produces results fully consistent with the method used in Fig.~\ref{fig:becfrac}.}.
In Fig.~\ref{fig:varyNB}, we show the results for four different values of the K atom number.  The BEC fraction in (a) decreases for a reduced number of bosons, but this can be fully attributed to the reduced critical temperature. The relative temperature in (b) shows a significant decrease for the lowest number of bosons.

Our results show that a reduction of the number of K atoms slightly improves the cooling performance of the Li Fermi gas. We interpret this observation as a consequence of the weak additional heat load associated with the bosons, which has to be removed by the evaporative cooling process. However, we do not observe any significant effect on the temperature of the Fermi sea if the number of K atoms stays below 12,000, which corresponds to about 3.0\% of the total number of $^6$Li atoms. The lowest temperature that we have observed in these measurements corresponds to $T/T_F \approx 0.06$.

\subsection{Thermalization and heating dynamics}
\label{ssec:therm}

\begin{figure}
\includegraphics[clip,width=1.00\columnwidth]{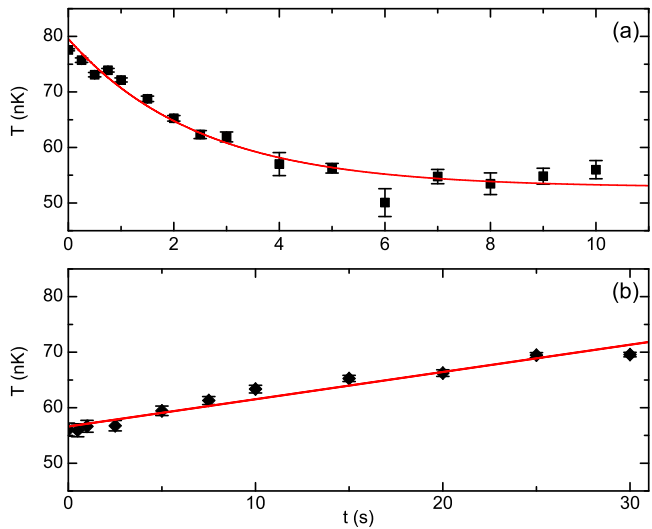}
\caption{Thermalization and heating dynamics with and without the Fermi sea. (a)  The data points show the measured temperature evolution of the bosonic $^{41}$K cloud [$N_B = 1.6(2)\times 10^4$] after evaporative cooling of the $^6$Li Fermi sea. The solid curve is an exponential fit, yielding a relaxation-time constant of $2.5(5)\,$s. (b) The data points display the measured temperature increase of the $^{41}$K cloud after full removal of the $^6$Li coolant. The linear fit (solid line) yields a heating rate of 0.49(4) nK/s. The error bars represent the temperature uncertainties as derived from the fit errors of the condensate fraction. }
\label{fig:thermalize}
\end{figure}

A central assumption underlying our paper is thermal equilibrium between the boson ''thermometer"and the Fermi sea. In order to test the validity of this assumption we have investigated the thermalization dynamics and residual heating effects that may affect our results. In all experiments discussed before, a hold time of 10\,s was introduced between the end of the evaporation ramp and the temperature measurement. We now present measurements on the temperature evolution during this hold time at a constant trap power of $P = 75$\,mW, again based on the detection of the condensate fraction. Figure \ref{fig:thermalize}(a) shows how the temperature drops from about 78\,nK right after the evaporation ramp to its equilibrium value of 53\,nK. An exponential fit yields a thermalization time scale of 2.5(5)\,s, which is short compared with the total hold time of 10\,s. This ensures that the K cloud reaches its equilibrium temperature with negligible deviations well below 1\,nK.

The thermalization time can be estimated from our experimental parameters, using the approximation
\begin{equation}
\frac{1}{\tau} = 2 \cdot \frac{3T}{2T_F} \cdot \frac{\xi}{3} \cdot n_F \sigma v_F \, ,
\label{eq:relax}
\end{equation}
which is a product of four factors. The prefactor of $2$ accounts for the two different spin states in the Fermi sea. The factor $3 T/(2 T_F) \approx 0.1$ results from the Pauli blocking of collisions \footnote{We approximate this effect by assuming that only the fraction of Li atoms with energies in an interval between $k_B (T_F-T/2)$ and  $k_B (T_F+T/2)$ is thermally active. The factor 3/2 results from the number of states, which increases $\propto E^{3/2}$ 
for the approximately homogeneous environment sampled by the bosons in the trap center.}. The third factor $3/\xi$  estimates the number of elastic collisions needed for thermalization,  with $\xi=4 m_{\rm K} m_{\rm Li}/(m_{\rm K}+m_{\rm Li})^2 \approx 0.45$ for the specific mass ratio of our mixture \cite{Mudrich2002scw}. 
The last factor represents the elastic collision rate in the limit of relative velocities dominated by the light atoms at the top of the Fermi sea, with the corresponding Fermi velocity $v_F = \sqrt{2 E_F /m} \approx 44$\,mm/s. The cross section for elastic collisions between $^6$Li and $^{41}$K atoms is $\sigma \approx 1.3 \times 10^{-16}$\,m$^2$ \cite{Hannapriv}. This results in a relaxation time of $\tau \approx 4.5$\,s, which is larger than the observed value, but still within the errors of the simple estimation used.

We have also checked the influence of a possible effect of residual heating of the K cloud, which may be induced by trap light fluctuations. This heat would have to be removed by thermal contact with the coolant, i.e.\ the Fermi sea of $^6$Li atoms, and the corresponding heat flow would require a temperature difference between the two components.  We have experimentally investigated the heating of the K cloud after removing the Li atoms from the trap, by application of a magnetic gradient, and observed the temperature evolution over 30\,s. Figure \ref{fig:thermalize}(b) reveals a very weak temperature increase, with a slope corresponding to a heating rate of $\gamma_{\rm heat} \approx 0.5$\,nK/s. Here, for simplicity, we assume a linear heating model. With the relaxation time $\tau \approx 2.5\,$s discussed before, we obtain a temperature difference of $\gamma_{\rm heat} \tau \approx 1.2\,$nK, which is negligibly small. In additional experiments, we have investigated heating in our detection trap, see Sec. \ref{ssec:detect}, and found an effect of less than 2\,nK/s within the 4\,s when the atoms are kept in this trap.

\subsection{Uncertainities}
\label{ssec:discuss}
Our thermometry approach is based on Eq.~(\ref{eq:basicideax}) to determine the relative temperature $T/T_F$. The underlying model relies on the harmonic approximation of the trap potential, and we estimate that anharmonicity effects on $T_F$ do not exceed a few percent. 

The model also assumes the bosonic probe to be a weakly interacting gas, which is well fulfilled. We have checked that we are not near any intraspecies or interspecies Feshbach resonances. Furthermore, the intraspecies background scattering length of $^{41}$K is about +60\,$a_0$ \cite{DErrico2007fri, Patel2014frm} and the background scattering length between $^6$Li and $^{41}$K is also about +60\,$a_0$ \cite{Hannapriv}.  This means that for the reference power ($P=75$\,mW) and $N_B \approx 1.2\times 10^4$, the chemical potential of the bosons corresponds to $\sim16$\,nK. The peak number density of the fermions is about 26 times smaller than the one of the bosons. The mean field of the fermions as seen by the bosons is very small, only $\sim2.3$\,nK. The correction to the boson trap frequencies caused by the fermion mean field, is on the order of $10^{-3}$\,, which is negligibly small. The mean field of the bosons on the fermions corresponds to $\sim64\,$nK, which is much smaller than the Fermi energy of about $710\,$nK. 

In addition to these model assumptions, the experimental determination of the temperature is subject to four main error sources. First of all, there are the statistical measurement uncertainities. These come from the analysis of the time-of-flight images and give uncertainties of a few percent in both the atom number and the determination of the condensate fraction. 

A second source that influences the measured values of $T/T_F$ are calibration errors. For the atom number determination, we estimate calibration uncertainities of 15\% for both species. This results in a systematic scaling uncertainty in $T/T_F$ of $\pm7$\%. Another systematic error source is the trap frequency ratio, which slightly changes if the trap does not exactly fulfill the magic levitation condition. However, the effect on $T/T_F$ for the range of powers used in our experiments is negligibly small.  

Thirdly, the thermalization between the two species may be imperfect, owing to heating in combination with weak thermal coupling. We estimate that the corresponding temperature difference stays below 2nK, which results in an effect below 3\% on the relative temperature.

Furthermore, as a fourth error source, we observed a slight heating effect during the transfer into the detection trap, which may also affect the temperature by a few percent at most. We are confident that other heating sources are very weak and can be safely neglected. All these residual heating effects may somewhat increase the temperature of the bosonic probe atoms, and may thus lead to an overestimation of the temperature, but not by more than 10\%.

Taking all statistical and systematic uncertainties into account, we can report a lowest observed temperature of $T/T_F = 0.059(5)$~\footnote{The reported temperature is the mean value of the  red circles, blue triangles, and green diamonds in Fig.~\ref{fig:varyNB}, in the range of $68-80$\,mW.}. The true temperature of the Fermi gas may even be slightly below this value (about 5\%) because of residual heating directly affecting the thermometry atoms.

\section{Conclusion}
We have thoroughly investigated a conceptually simple and accurate method for determining the temperature of a deeply degenerate Fermi gas. Our method essentially relies on detecting the condensate fraction of a second, weakly interacting bosonic species that is thermalized with the Fermi sea. High accuracy of the temperature measurements can be achieved, since the relevant trap frequency ratio can be very well determined and uncertainties in the atom number only weakly influence the results. 

We have employed the method in experiments on a spin mixture of $^6$Li, where we have used a small sample of $^{41}$K bosons as the probe. The large mass ratio and a large number ratio have enabled us to measure the temperature in the range of 0.03 to 0.1 $T_F$, which is right in the regime of state-of-the art cooling experiments. We have investigated the final stage of deep evaporative cooling and we have observed that the deepest degeneracy of the Fermi gas, with $T/T_F = 0.059(5)$, is achieved when the evaporation is stopped just before the onset of spilling. We found the temperature not to be affected by the presence of the probe atoms if the number of K atoms stays below 3.0\% of the number of Li atoms. The K atoms then represent impurities in a Fermi sea. 

Our thermometry method provides us with a powerful tool to further optimize the cooling. For optimization, we can improve the starting conditions for evaporation by implementing a sub-Doppler cooling stage \cite{Burchianti2014eap, Sievers2015ssd} and we can optimize the evaporation sequence by variation of the magnetic field, the trap configuration and the details of the ramp. With sensitive and accurate thermometry, it will be very interesting to investigate the practical and fundamental limitations of the cooling process. Under our present conditions, we may be limited by residual trap light fluctuations~\cite{Savard1997lni} or other sources of noise in the experiment or by inelastic losses in combination with the hole heating effect~\cite{timmermans2001dfg}. 

For the interaction parameter of $1/(k_F a) \approx -1.6$, as chosen in our experiments, the predicted critical temperature for superfluidity is $\sim0.03\,T_F $ \cite{Bloch2008mbp, Haussmann2007tot}. Thus, even for our lowest temperatures, the Li spin mixture is not superfluid.  However, the system is stable enough at resonant interaction conditions  \cite{Spiegelhalder2009cso}, so that the realization of a mass-imbalanced Bose-Fermi double superfluid, as already demonstrated in Ref.~\cite{Yao2016ooc}, would be straightforward. 
Thermometry on the bosons could be performed in a wide range of the BEC-BCS crossover, as long as the thermalization time stays much shorter than the timescale of inelastic losses~\footnote{Inelastic decay of $^{41}$K is observed predominantly on the BEC side of the $^{6}$Li Feshbach resonance, similar to what was observed on a strongly interacting $^{40}$K-$^{6}$Li mixture~\cite{Spiegelhalder2009cso}}. While the BEC side may be problematic~\cite{Spiegelhalder2009cso}, the method would work well in the unitary case and on the BCS side.

The implementation of the presented thermometry method should be straightforward for other Bose-Fermi mixtures. Extreme mass ratios~\cite{Pires2014ooe, Tung2014gso, Roy2016tem, Konishi2016cso} are of particular interest for pushing the accessible regime further down to temperatures on the order of 0.01\,$T_F$. However, at such ultralow temperatures, the larger number of collisions required for thermalization and the increasing Pauli blocking effect will increase the thermalization time, which will make it more difficult to reach thermal equilibrium on a realistic experimental time scale. This may be compensated for by larger interspecies collision cross sections or higher number densities. Our paper shows how the conditions can be optimized for specific mixtures, including the role of optical polarizabilities, magnetic moments, magnetic levitiation for trapping, and the effect of interspecies collisions.

In our future work, we are particularly interested in the deep cooling of the Fermi sea. This reduces thermal decoherence effects as observed in studies of impurities coupled to the Fermi sea \cite{Cetina2015doi} and opens up the possibility of observing new phenomena \cite{Cetina2016umb}, such as multiple particle-hole excitations and the onset of the orthogonality catastrophe \cite{Knap2012tdi}. Moreover, we are interested in the collective zero-temperature dynamics of bosonic impurities in the Fermi sea close to an interspecies Feshbach resonance \cite{Lous2017psi}.

\begin{acknowledgments}
We thank J. Walraven for stimulating discussions, Y. Colombe for fiber fusing, L. Reichs\"ollner for support regarding the atomic-beam characterization, the Dy-K team for general discussions, and M. Cetina for contributions in the early stage of the $^{41}$K implementation. We also thank F. Lehmann, E. Kirilov and R. Onofrio for comments on the paper. We acknowledge support by the Austrian Science Fund FWF within the Spezialforschungsbereich FoQuS (F4004-N23) and within the Doktoratskolleg ALM (W1259-N27). 
\end{acknowledgments}

\section{Appendix: `Magic' levitation trap}
\label{app:magic}

We refer to a `magic' levitation trap as an optical dipole trap for two different species, in which the corresponding potential depths and trap frequencies  maintain a constant ratio at any optical power applied. In optical dipole trapping experiments, one often has to deal with the effect of gravity. Two species in the same trap are in general affected differently, in particular in the case of largely different masses or different optical polarizabilities. The tilted potentials usually give a different reduction of the effective trap depth as compared to the depth of the optical potentials. During evaporative cooling this often leads to a much faster reduction of the potential depth for the heavier species than for the lighter one, which may pose a severe limitation to the whole cooling process. Magnetic levitation \cite{Anderson1995oob,Han2001lac,Weber2003bec} offers a solution to this problem and allows one to realize conditions, where the combined effect of gravity and levitation results in the same effect on the total shape of the potential.

\begin{figure}
\includegraphics[clip,width=0.75\columnwidth]{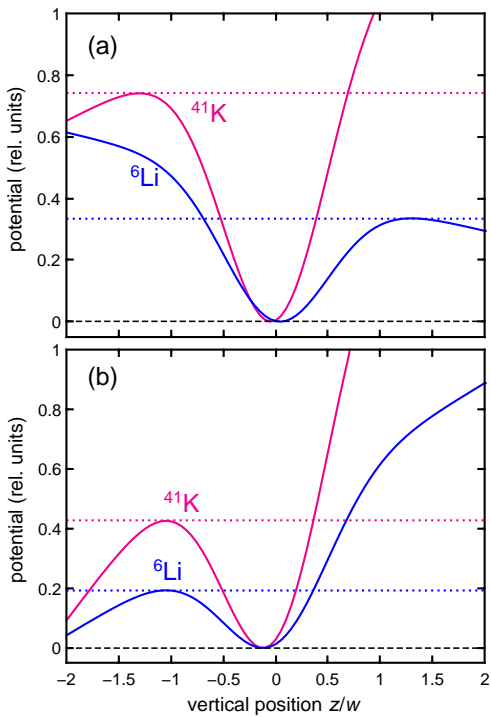}
\caption{(Color online) Illustration of magic levitation for $^6$Li and $^{41}$K. All potentials are normalized to the optical potential depth of K, and zero potential refers to the trapping potential minima. The combined magnetic and gravitational forces cause a trap depth reduction, as indicated by the horizontal dotted lines. For two distinct values of the magnetic gradient, see Eq.~(\ref{eq:A2}), the ratio of the resulting trap depths for K and Li remains constant and corresponds to the ratio of optical polarizabilities ($\alpha_{\rm K}/\alpha_{\rm Li} \approx 2.2$).
(a) With the magnetic gradient set to $B'_-$, K is partially levitated, while Li is overlevitated. The relative tilt has the same magnitude, but opposite sign. (b) With the gradient set to $B'_+$, the magnetic force effectively increases the effect of gravity for both species, resulting in a tilt in the same direction.}
\label{fig:A}
\end{figure}

The magic gradient can be derived from the condition that the combined magnetic and gravitational force is the same for both traps, if considered relative to the optical potential, the depth of which in turn is proportional to the optical polarizabilities. The condition reads
\begin{equation}
\frac{\mu_1 B_{\pm}' - m_1 g}{\alpha_1} = \pm \, \frac{\mu_2 B_{\pm}' - m_2 g}{\alpha_2} \, ,
\label{eq:A1}
\end{equation}
where $m_i$, $\mu_i$, and $\alpha_i$ represent the different masses, magnetic moments, and optical polarizabilities of the two species, respectively.  The lower sign refers to the
situation  illustrated in Fig.~\ref{fig:A}(a), where the trapping potentials are tilted in the opposite direction. The upper sign corresponds to the situation, where both potentials are tilted in the same direction. Solving Eq.~(\ref{eq:A1}) yields the two corresponding magnetic gradients
\begin{equation}
B'_{\pm} = \frac{\alpha_2 m_1 \mp \alpha_1 m_2}{\alpha_2 \mu_1 \mp \alpha_1 \mu_2} \, g \, .
\label{eq:A2}
\end{equation}
The solution $B'_-$ means partial levitation for one species and overlevitation for the other one, so that the tilts have opposite signs. The other solution ($B'_+$) corresponds to the same direction of the tilt for both species. The application of $B'_-$ causes a separation of the trap centers, similar to the gravitational sag effect. In contrast, $B'_+$ does not cause such a spatial shift, but it may imply much stronger tilts. The optimum solution for an experiment depends on the specific situation. 

For our situation of optically trapped $^6$Li and $^{41}$K at high magnetic bias fields ($\mu_1 = \mu_2 \approx \mu_B$), we obtain a magic levitation gradient of $B'_- = 2.97\,$G/cm, corresponding to a partial levitation of 41.3\% for K and an overlevitation of 281\% for Li. The small spatial separation of the trap centers
 is irrelevant for our application. For the experimental power range we use, the separation between the trapcenters of the two species lies between 12 and 28 \% of the optical beam waist.
Note that the other solution ($B'_+ = -4.02\,$G/cm) does not correspond to levitation, but to an effective increase of the gravitational effect for both species. As described in Sec.~\ref{ssec:trap}, we realize a situation close to the magic levitation gradient $B'_-$.

\bibliographystyle{apsrev4-1}
\bibliography{ultracold,thermometryNewrefs}

\end{document}